\documentclass[a4paper,11pt]{article}
\usepackage{graphicx}
\usepackage{subfig}
\usepackage{here}
\usepackage{color}
\usepackage{amsmath}
\usepackage{hyperref}
\usepackage{url}
\newcommand{\siki}[1]{Eq.~(\ref{eq:#1})}

\newcommand{\ep}{\varepsilon}

\begin{document}

\title{
Analytic and numerical approaches for depictive 3-loop integrals using sector decomposition}

\author{Elise de Doncker${}^{\mathrm{a}}$, 
Tadashi Ishikawa${}^{\mathrm{b}}$, 
Kiyoshi Kato${}^{\mathrm{c,}}$\footnote{e-mail~:~{\tt katok@kute.tokyo}} , 
Fukuko Yuasa${}^{\mathrm{b}}$ \\
{ \small ${}^{\mathrm{a}}$Western Michigan University, Kalamazoo MI 49008, USA, } \\
{ \small ${}^{\mathrm{b}}$High Energy Accelerator Research Organization (KEK), 1-1 Oho Tsukuba,}  \\
{ \small Ibaraki 305-0801, Japan,} \\
{ \small ${}^{\mathrm{c}}$Kogakuin University, Nishi-Shinjuku 1-24, Shinjuku, Tokyo 163-8677, Japan} }

\date{}

\maketitle

\begin{abstract}
Four 3-loop two-point functions are studied analytically and numerically
using a simplified sector decomposition method.
The coefficients of the ultraviolet divergent part are determined analytically,
and those of the finite part are computed numerically.
The energy dependence of the integrals is shown explicitly, and a discussion of its behavior is presented.
\end{abstract}


\section{Introduction}

As high-precision experimental data accumulates, it becomes 
increasingly important to estimate the theoretical radiative correction 
that matches its accuracy. 
Additionally, by examining the contribution of virtual particles 
included in the radiative correction, it would be possible to predict 
unknown particles. The evaluation of such radiative corrections 
requires the calculation of loop diagrams with masses as an essential element.

The sector decomposition (SD) method~\cite{nakanishi71,binoth00,binoth04,binoth04a,heinrich08a,heinrich21} is well-established 
for calculating loop integrals. In this paper, we use a simple form of 
the SD method. Conventionally, the number of regions 
increases depending on the number of particles and the degree of the loops, 
which has been a problem. Therefore, various techniques~\cite{bogner08,bogner08a,SmirnovTentyukov09,smirnov09,kaneko10} have been used to 
suppress this increase. However, with the advances in computer 
capabilities, it has been found that the  method can achieve 
precise calculations up to a certain level by combining it with automatic processing of the integration. 
From the basic formula derived by this method, 
it is possible to obtain a specific expression of the coefficients that represent the ultraviolet (UV) divergence analytically, 
and numerical calculations with sufficient accuracy can be performed 
even for the non-divergent terms.

We have developed fully numerical approaches~\cite{jocs11,cpc16,ddacat17}  
using non-adaptive or adaptive integration 
with transformations and/or extrapolation 
to manage the singular behavior
of loop integrals.
The method consists of an accurate integration and estimation of the coefficients. The former includes
the numerical integration techniques, 
such as adaptive integration~\cite{pi83,parintweb,csci19}, the double-exponential formula~\cite{mori78}, 
and Quasi-Monte Carlo integration based on lattice rules~\cite{sloan}, 
and the latter uses nonlinear extrapolation~\cite{shanks55,wynn56} or a  linear solver.
We have succeeded to manage the singularity from both
the UV-divergence and in the physical kinematics below and beyond the threshold in the squared momentum.
To achieve this, we use a double extrapolation technique and
have already shown some results for the 2-loop~\cite{talk-ACAT2022} and for the 3-loop order~\cite{talk-CCP2023}.

In the following sections, as a concrete example of this method,
we provide results  
for several 3-loop diagrams in Fig.~\ref{fig:diags} that are of a form we call 
``complete''.  
We deal with the UV singularity by dimensional regularization, 
where the space-time dimension is set to $n=4-2\ep$. 
A 3-loop integral ${\bf I}$ is given by its Laurent series in $\ep$:
\begin{equation}
{\bf I}=
\frac{{\bf C}_{-3}}{\ep^3} + \frac{{\bf C}_{-2}}{\ep^2} 
+ \frac{{\bf C}_{-1}}{\ep} + {\bf C}_0 + \cdots \,.
\label{eq:expandzero}
\end{equation}
As referenced in Section~2, the Loop (I), sometimes called banana diagram, 
is studied in the literature~\cite{bloch15,broedel22,kreimer23,weinzierl22c}
based on a general framework.
The Loop (I), (II) and (IV) diagrams are studied in ~\cite{martin23,martin22,bauberger20} 
using integration-by-parts (IBP) relations for the generic mass case.

Further in Section~2, we give the formulation of our calculation approach
and analytical results for the divergent part, i.e., 
for ${\bf C}_{-3}, {\bf C}_{-2}$ and ${\bf C}_{-1}$. 
In Section~3, we describe the numerical methods 
and the results by numerical computation for the coefficients up to and including ${\bf C}_0$.
Finally, we summarize and provide future prospects in Section~4.

\section{Formalism and analytic results}

We define the scalar integral with numerator $1$  as
\begin{equation}
{\bf I}= \prod_{k=1}^L \int \frac{d^n \ell_k}{(2\pi)^n} \frac{1}{D_1\ldots D_N}
\end{equation}
where $D_j=p_j^2-m_j^2$ is the inverse propagator of the $j$-th line,
$N$ is the number of internal lines and $L$ is the number of loops.

Subsequently we consider 3-loop two-point functions, so that $L=3$.
The external momentum that flows into the diagram is $P$ and we denote $s=P^2$.
Using the Nakanishi-Cvitanovi\'c-Kinoshita 
formalism~\cite{nakanishi57,kinoshita74A,kinoshita74B,kinoshita74C}, the integral is given by
\begin{equation}
{\bf I}=(-1)^N\frac{\Gamma(N-3n/2)}{(4\pi)^{3n/2}}
\int \prod dx_{j} 
\frac{\delta(1-\sum x_j)}{U^{n/2} V^{N-3n/2}} \ .
\end{equation}
\begin{equation}
V=M^2-s\frac{W}{U} , \qquad M^2=\sum m_j^2 x_j - i \rho
\label{eq:integuv}
\end{equation}
where $U$ ($W$) is a sum of monomials and each monomial is a product 
of three (four) different $x$'s with coefficient $1$.
Here, $i\rho$, the infinitesimal imaginary part of mass, is clearly expressed
to be used in the numerical study of the next section. 
Factors are separated from the integral as follows:

\begin{equation}
{\bf I}=(-1)^N\left(\frac{1}{(4\pi)^{2-\ep}}\right)^3 \Gamma(1+3\ep)\times \tilde{I} , \qquad
\tilde{I} =\frac{\Gamma(N-6+3\ep)}{\Gamma(1+3\ep)} \times I 
\end{equation}

\begin{equation}
I=\int d{\bf \Gamma}
\frac{V^{6-N-3\ep}}{U^{2-\ep} }, \qquad
d{\bf \Gamma} = \prod_{j=1}^N dx_{j} ~\delta(1-\sum x_j)
\label{eq:integzero}
\end{equation}
The UV singularity appears in the Gamma function if $N\le 6$, and in the integral when
$U=0$, so that the UV singularity of $I$ is $O(\ep^{-2})$ or less:
\begin{equation}
I= \frac{C_{-2}}{\ep^2} + \frac{C_{-1}}{\ep^1} + C_0 + C_1 \ep  + \cdots
\label{eq:expandone}
\end{equation}
and the ${\bf C}$'s in \siki{expandzero} can be calculated from the $C_j$.

\begin{figure}[htb]
\begin{center}
\includegraphics[width=0.99\linewidth]{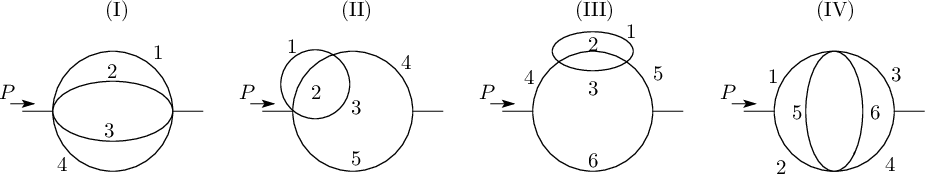}
\end{center}
\caption{3-loop diagrams}
\label{fig:diags}
\end{figure}

We consider the (I), (II), (III), (IV) 3-loop diagrams in Fig.~\ref{fig:diags} where $N=4, 5, 6, 6.$ 
The Feynman parameter $x_j$ is assigned to the
line $j$ in the figure.
The $U, W$ functions are as given below.  For the Feynman parameters,
we use the notation $x_{ab\cdots}=x_a + x_b + \cdots$. 

\begin{equation}
\begin{array}{ll}
\mathrm{Loop~(I)} &
U= x_1x_2x_3 + x_1x_2x_4 + x_1x_3x_4 + x_2x_3x_4 ,\\
 & W =  x_1x_2x_3x_4 . \\
\mathrm{Loop~(II)} &
U=x_1x_2x_3+(x_1x_2 +x_1x_3+x_2x_3)x_{45} , \\
 & W=(x_{12}x_3x_4+x_1x_2x_{34})x_5 . \\
\mathrm{Loop~(III)} &
U = x_1x_2x_3 + (x_1x_2 + x_1x_3 + x_2x_3)x_{456} , \\
 & W = (x_1x_2x_3 + (x_1x_2 + x_1x_3 + x_2x_3)x_{45})x_6 .  \\
\mathrm{Loop~(IV)} &
U = x_{12}x_{34}x_{56}+(x_{12}+x_{34})x_5x_6 , \\
 & W=(x_1x_2x_3 + x_1x_2x_4 + x_1x_3x_4 + x_2x_3x_4) x_{56} +x_{13}x_{24}x_5x_6 . \\
\end{array} 
\end{equation}

\noindent
We introduce the following variable transformations. Here $\bar{a}=1-a$.

\begin{equation}
\begin{array}{llllll}
\mathrm{Loop~(I)} & 
x_1=y_1, & x_2=y_2, & x_3=y_3, & x_4=y_4 . \\
\mathrm{Loop~(II)} &
x_1=y_1, & x_2=y_2, & x_3=y_3, & x_4=\bar{w} y_4 , & x_5=w y_4 .  \\
\mathrm{Loop~(III)} &
x_1=y_1, & x_2=y_2, & y_3=y_3 , \\
& x_4=y_4 w \bar{z}, & 
x_5=y_4 w z, & 
x_6=y_4 \bar{w} . \\
\mathrm{Loop~~(IV)} & 
x_1=y_1 \bar{w}_1, & 
x_2=y_1 w_1, & 
x_3=y_2 \bar{w}_2, & 
x_4=y_2 w_2 , \\
 & x_5=y_3, & x_6=y_4 . \\
\end{array}
\end{equation}

\noindent
After the transformation, all $U$ functions are of the form
\begin{equation}
U= y_1y_2y_3 + y_1y_2y_4 + y_1y_3y_4 + y_2y_3y_4 .
\end{equation}
We consider the $U$ function being ``complete'' in the sense that all
possible monomials in $y_1, y_2, y_3$, and $y_4$ are present.
A consequence of this property is described later.
The $W$ function is also expressed in terms of the $y$'s and additional
variables ($z, w$).

The phase space can be written as
\begin{equation}
\begin{array}{ll}
\mathrm{Loop~(I)} & 
d{\bf \Gamma} = d \tilde{\Gamma} \\
\mathrm{Loop~(II)} &
d{\bf \Gamma} = dw\, d \tilde{\Gamma}\, y_4 \\ 
\mathrm{Loop~(III)} &
d{\bf \Gamma} = w\, dw\, dz\,  d \tilde{\Gamma}\, y_4^2 \\
\mathrm{Loop~(IV)} & 
d{\bf \Gamma} =  dw_1\, dw_2\, d \tilde{\Gamma}\, y_1y_2 \\
\end{array}
\end{equation}
where
\begin{equation}
d\tilde{\Gamma}=  dy_1\, dy_2\, dy_3\, dy_4\, \delta(1-\sum y_j) 
\end{equation}
and 
\begin{equation}
R=\{(y_1, y_2, y_3, y_4)~|~\sum_j y_j=1 \}
\end{equation}
is the full phase space in the $y$'s.

Hereafter, we assume that all masses are the same, 
\begin{equation}
m=m_j \quad (j=1,\ldots,N). 
\end{equation}
Extension to the general case is possible.
In Loop (III), if $m_4=m_5$, the $z$-dependence disappears.

The calculation of \siki{integzero}
is done by the following procedure.
The region $R$ is divided into regions $R(klm)$ 
in the following three steps 
where $k, l, m, n$ is a permutation\ of $1, 2, 3, 4$.

\noindent \underline{First step}: 
The region $R$ is divided into four regions as $R=\cup_j R(j)$ where
$R(k) = \{(y_1, y_2, y_3, y_4)$ $\in R~|~ y_k = \mathrm{max}_j \{y_j\} \}$.
In $R(k)$ we perform the transformation
$y_j= s_k t_j,\, t_k=1$ and
we integrate over $s_k$ to eliminate the delta-function.
Then $R(k)$ is the 3-dimensional unit cube of $(t_l, t_m, t_n)$.
The phase space is
\begin{equation}
d \tilde{\Gamma} = s_k^4 dt_l dt_m dt_n 
\end{equation}
and $s_k^{-1}= \sum_{j=1}^{4} t_j$.

\noindent \underline{Second step}: 
The region $R(k)$ is divided into three regions as $R(k)=\cup_j R(kj)$ where
$R(kl) = \{(t_l, t_m, t_n) \in R(k)~|~ t_l = \mathrm{max}_j \{t_j\} \}$.
In $R(kl)$ we perform the transformation
$t_l=t_l,\,  t_m=t_l u_m,\,  t_n=t_l u_n$. 
The phase space is $ dt_l dt_m dt_n= t_l^2 d t_l d u_m d u_n$.

\noindent \underline{Third step}: 
The region $R(kl)$ is divided into two regions as $R(kl)=\cup_j R(klj)$ where
$R(klm) = \{(t_l, t_lu_m, t_lu_n)\in R(kl)~|~ u_m = \mathrm{max}_j \{u_j\} \}$.
In $R(klm)$ we perform the transformation
$ u_m= u_m,\,  u_n=u_m v_n$. 
The phase space is $ dt_l dt_m dt_n= t_l^2 u_m d t_l d u_m d v_n$.

In summary, in the region $R(klm)$, the $y$-variables are transformed as
\begin{equation}
y_k= s_k, \quad
y_l= s_k t_l, \quad
y_m= s_k t_l u_m, \quad
y_n= s_k t_l u_m v_n 
\label{eq:strans}
\end{equation}
and the $U,\,V$ functions are
\begin{equation}
U = s_k^3 t_l^2 u_m f, \qquad
W = s_k^4 t_l^2 u_m q, \quad
V/s_k=G_k = m^2 s_k^{-1} - s \frac{q}{f} -i \rho \, ,
\end{equation}
where the $f$ and $q$ functions are introduced.

The integral in the region $R(klm)$ is denoted as $I(klm)$ and 
 $I$ is expressed as the sum of the integrals $I(klm)$.
Hereafter, we omit the subscript from $t, u, v, G$ when it is redundant.
\begin{equation}
\begin{array}{lll} 
\mathrm{Loop~(I)} &
I(klm) = {\displaystyle \int d\Gamma
 \frac{h}{t^{2-2\ep}u^{1-\ep}}H }, & h=1 ,    \\
 { } & { } & { } \\
\mathrm{Loop~(II)} &
I(klm) = {\displaystyle  \int_0^1 dw \, \int d\Gamma
 \frac{h}{t^{2-2\ep}u^{1-\ep}}H }, & h=  y_4/s_k  , \\
 { } & { } & { } \\
\mathrm{Loop~(III)} & 
I(klm) = {\displaystyle  \int_0^1 w\,dw \, \int d\Gamma
 \frac{h}{t^{2-2\ep}u^{1-\ep}}H }, & h=  y_4^2/s_k^2 ,   \\
 { } & { } & { } \\
\mathrm{Loop~(IV)} &
I(klm) = {\displaystyle  \int_0^1 dw_1 \int_0^1 dw_2\,  \int d\Gamma
 \frac{h}{t^{2-2\ep}u^{1-\ep}}H }, & h=y_1y_2/s_k^2 \, .   \\
\end{array}
\label{eq:integfin}
\end{equation}

\begin{equation}
 \int d\Gamma =  \int_0^1 dt \int_0^1 du \int_0^1 dv \, , \qquad
H=\frac{G^{6-N-3\ep}}{f^{2-\ep} } .
\end{equation}
\begin{equation}
f=  1+v+uv+tuv, \qquad 
G = m^2 (1 + t ( 1+  u (1+ v) ) ) - s \frac{q}{f}-i \rho  .
\label{eq:fG}
\end{equation}
\siki{integfin} is our basic formula to calculate the
loop integral analytically and numerically.
It should also be noted that $s_k$ does not occur in $h$ or $q$.
The form of the $f$ function after the three
transformation steps is  shown in \siki{fG} to be that of a regular function, and the UV singularity 
from the $t$- and $u$-integrals is under control as the explicit factor in \siki{integfin}.
Also,  the $f$ function is common in all regions by the ``completeness''.
If $U$ is not ``complete'', we need another transformation step in
several regions and the $f$ function is determined in each region.

Although there are 24 regions $R(klm)$, since we consider the case of equal masses,
the integral $I$ is expressed as follows using the symmetry of the diagram:

\begin{equation}
\begin{array}{ll}
\mathrm{Loop~(I)} &
I= 24 I(123) \\
\mathrm{Loop~(II)} &
I= 6( I(123) + I(124) + I(142) + I(412) ) \\
\mathrm{Loop~(III)} & 
I= 6( I(123) + I(124) + I(142) + I(412) ) \\
\mathrm{Loop~(IV)} &
I= 4( I(123) + I(132) + I(312) + I(341) + I(314) + I(134) ) 
\end{array}
\end{equation}
In each region, $h$ and $q$ can be determined by \siki{strans}.

A procedure for the computation of $I(klm)$ is sketched in Appendices A and B.
We present the analytic results for the first three coefficients of $I$ 
in the Laurent series in $\ep$.



\vspace{5mm}
\noindent\underline{Loop (I)}

\[
I=\frac{1}{\ep^2} 12(m^2)^2 
-  \frac{1}{\ep} m^2\left(2s+8m^2+ 36 m^2 \log m^2 \right)
\]
\begin{equation}
+ \frac{1}{6}s^2 -(6\pi^2+48) (m^2)^2 
  +  m^2 \log m^2 \left( 6s + 24 m^2 + 54m^2 \log m^2 \right) 
   +O(\ep)
\label{eq:loopdone}
\end{equation}

\vspace{5mm}
\noindent\underline{Loop (II)}

\[
I = \frac{1}{\ep^2} 3m^2  
 + \frac{1}{\ep} 6m^2 \left( -\frac{1}{3} -\frac{3}{4}(j_s^{(1)}(s)+\log m^2) \right) 
 + \frac{3}{4}m^2(1+12 \log m^2) j_s^{(1)}(s) 
\]
\begin{equation}
 + \frac{9}{4}m^2 j_s^{(2)}(s)
 + \frac{1}{4}s + \frac{3}{4}m^2(7\log m^2 + 3 (\log m^2)^2) 
  - 6m^2( \frac{\pi^2}{4} + 2)
   +O(\ep)
\label{eq:loopdtwo}
\end{equation}

\vspace{5mm}
\noindent\underline{Loop (III)}

\begin{equation}
I = - \frac{1}{\ep} \, \frac{1}{4}(9 j_t^{(0)}(s) + 1 ) 
  + 6\left(-\frac{3}{16}+\frac{1}{8}j_s^{(1)}(s) +\left(\frac{3}{4}\log m^2-\frac{17}{16}\right)j_t^{(0)}(s)
  + \frac{3}{8}j_t^{(1)}(s) \right)
 +O(\ep)
\label{eq:loopdthr}
\end{equation}
In this loop instance, the $O(\ep^{-2})$ term happens to be vanishing.

\vspace{5mm}
\noindent\underline{Loop (IV)}

\begin{equation}
I = \frac{1}{\ep^2} 
+ \frac{1}{\ep} ( 1 - 3 j_s^{(1)}(s) )
+ \left( \frac{3}{2}j_s^{(2)}(s)  + 3 (j_s^{(1)}(s))^2 - 3 j_s^{(1)}(s) - \frac{\pi^2}{2} + 1 \right)
 +O(\ep)
\label{eq:analytic2j}
\end{equation}

\vspace{5mm}
In the above formulae, we use the following functions where
$D(x)=m^2-s x\bar{x}$.
\footnote{
Here, $j_t^{(0)}=(2 j_s^{(1)}+4-2\log m^2)/(4m^2-s)$ and
$j_t^{(1)}=m^2(j_s^{(2)}+4j_s^{(1)}-(\log m^2)^2)/(4m^2-s)$.
}

\begin{equation}
j_s^{(k)}(s) = \int_0^1 dx\, [\log D(x)]^k , \quad
j_t^{(k)}(s)=  \int_0^1 dx\, \frac{m^2 [\log D(x)]^k}{D(x)} . 
\end{equation}

The formulae in \siki{loopdone}, \siki{loopdtwo} and \siki{analytic2j} agree with those in ~\cite{martin23,martin22}.
The \siki{loopdthr} is obtained as the derivative of the formula in ~\cite{martin23,martin22}
corresponding to \siki{loopdtwo} by $m_4^2$.
In the above, we give the expression for the UV divergent terms, $C_{-2}, C_{-1}$ and $C_{0}$. 
The analytic evaluation of the fourth term is highly difficult and the calculation
of the $C_1$ term is studied by the numerical method in the next section.

\vspace{3mm}

In the present approach, we sometimes encounter a pseudo-threshold.
As an example, in the calculation of the $O(1)$ term of Loop (III), the
following term appears:
\begin{equation}
 \int_0^1 dx\,  \frac{(2m^2-sx\bar{x}) \log (2m^2-sx\bar{x})}{D(x)},
\end{equation}
which has the threshold at $s=8m^2$.
While such terms exist in several regions, in the sum they cancel each other out,
so that the threshold at $s=8m^2$ vanishes.
In the calculation of the $O(\ep)$ terms, we find more
complicated pseudo-thresholds.  
The numerical integration is done in each region, so that 
the appearance of a pseudo-threshold 
might affect the high-precision computation.
Further study will be required to clarify this behavior.

\section{Numerical results}
The numerical calculation of \siki{integfin} is done in the following way.

We consider the coefficients in \siki{expandone} and there are two approaches.
In the first,  $\ep$ no longer stands for an infinitesimal 
quantity but is a (small) finite value.
We prepare a descending positive-number sequence $\{\ep_j\}$ and
calculate the $I(klm)$ taking $\ep=\ep_j$ to obtain a value $I_j$.
From the sequence of values $\{I_j\}$, we estimate the $C's$ in \siki{expandone}
by a linear or nonlinear extrapolation method~\cite{cpp10,acat11,cpc16}.
If the singularity at $t=0$ or $u=0$ is of type $ \int_0^1 dx\,x^{-1+\ep}f(x)$
where $f$ is a non-singular function,
the calculation is just a numerical integration since $\ep$ is
a positive finite quantity.
If the behavior of the integrand is of type $\int_0^1 dx\,x^{-2+\ep}f(x)$, 
we use the treatment described in Appendix A. 
Another approach is, after the separation of the UV singularity of \siki{integfin}
as shown in Appendix B, by expanding the integrand in a Maclaurin series in $\ep$  
and calculating each term using numerical integration to obtain
the coefficients of \siki{expandone}.

Above the threshold, the imaginary part appears in $I$. In this case,
we no longer treat the $\rho$ in \siki{integuv} as infinitesimal.
We prepare a geometric sequence of numerical values ${\rho_j}$
and calculate the integrals $I(klm)$ with $\rho=\rho_j$ to obtain a value $I_j$.
Then we perform the estimation of $I$ as the limiting value for $\rho \rightarrow 0$. 
The extrapolations with respect to $\rho$ and $\ep$ can be combined 
to constitute a double extrapolation~\cite{talk-ACAT2022,talk-CCP2023}.
\begin{figure}[htb]
\begin{center}
\includegraphics[width=0.49\linewidth]{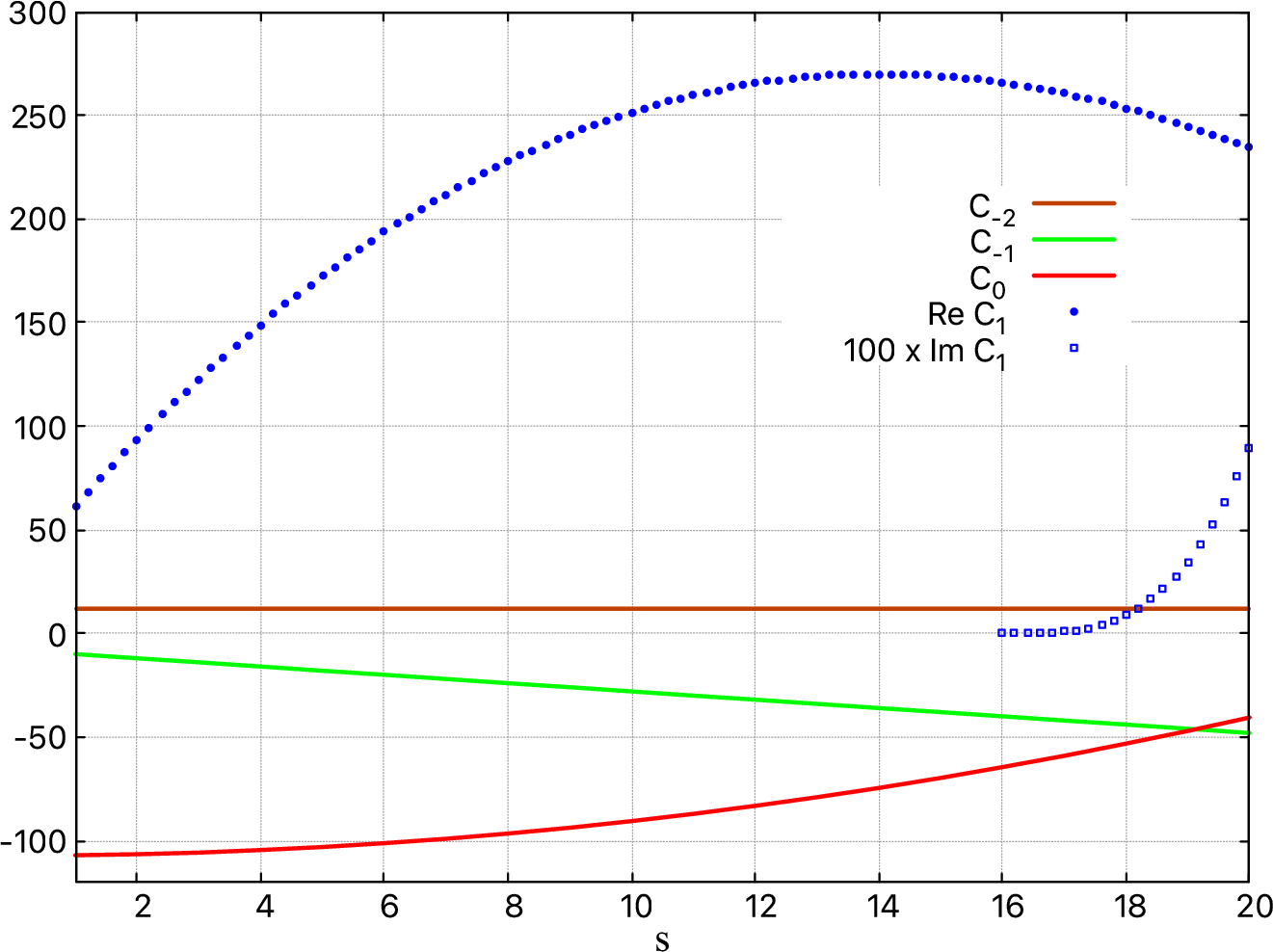}
\end{center}
\caption{Expansion coefficients of Loop (I), $C_{-2}$ (dark-orange solid line), $C_{-1}$ (green solid line), 
$C_0$ (red solid line) and $C_1$ (real part in blue dots and imaginary part in blue diamond boxes) as a function of $s$. The imaginary part of $C_1$  appears for $16m^2 < s$ and it is multiplied by 100 to fit in the plot.}
\label{fig:L-I}
\end{figure}

\begin{figure}[htb]
\begin{center}
\hspace*{0.3cm}
\subfloat[]{\includegraphics[width=0.49\linewidth]{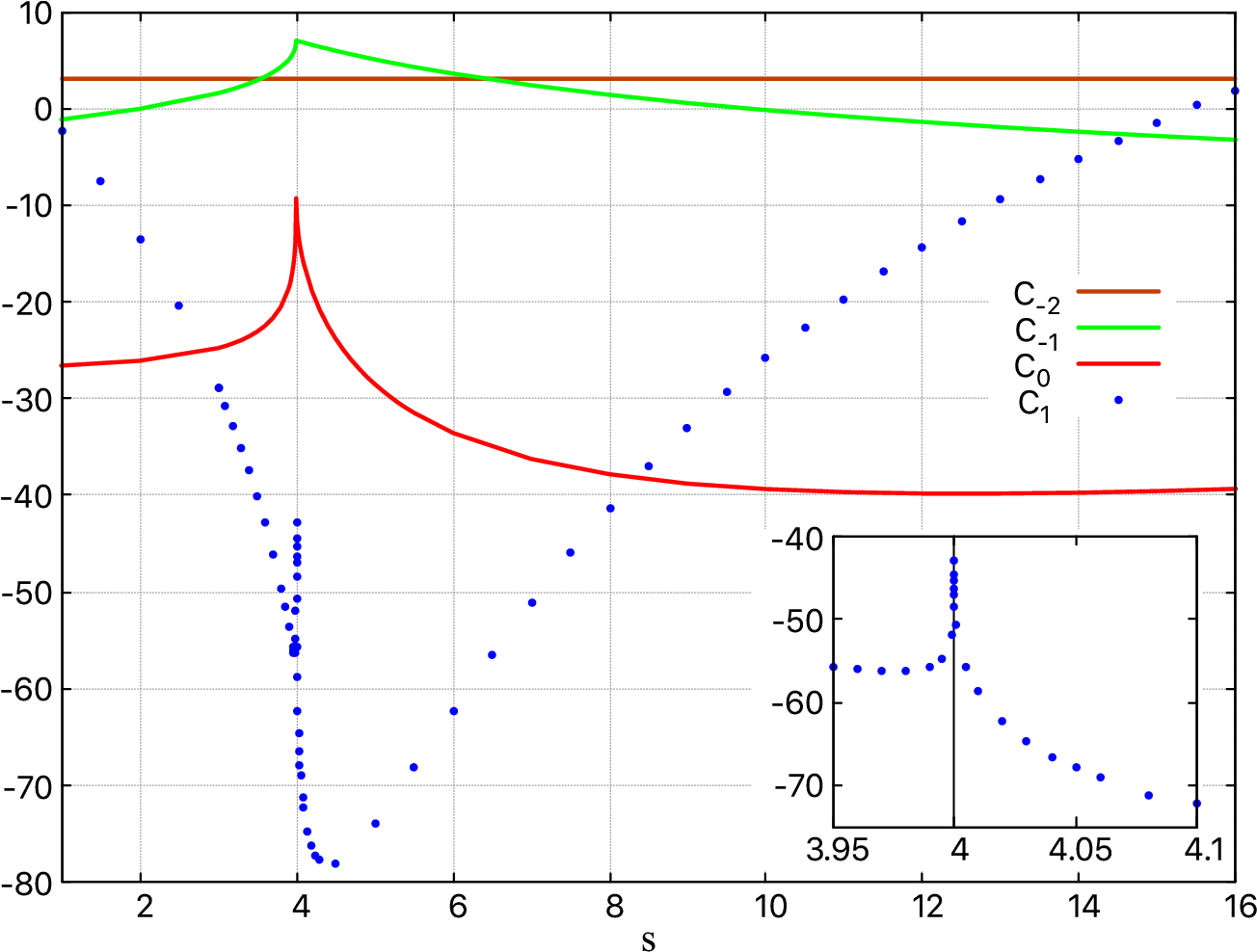}}
\hspace*{0.3cm}
\subfloat[]{\includegraphics[width=0.49\linewidth]{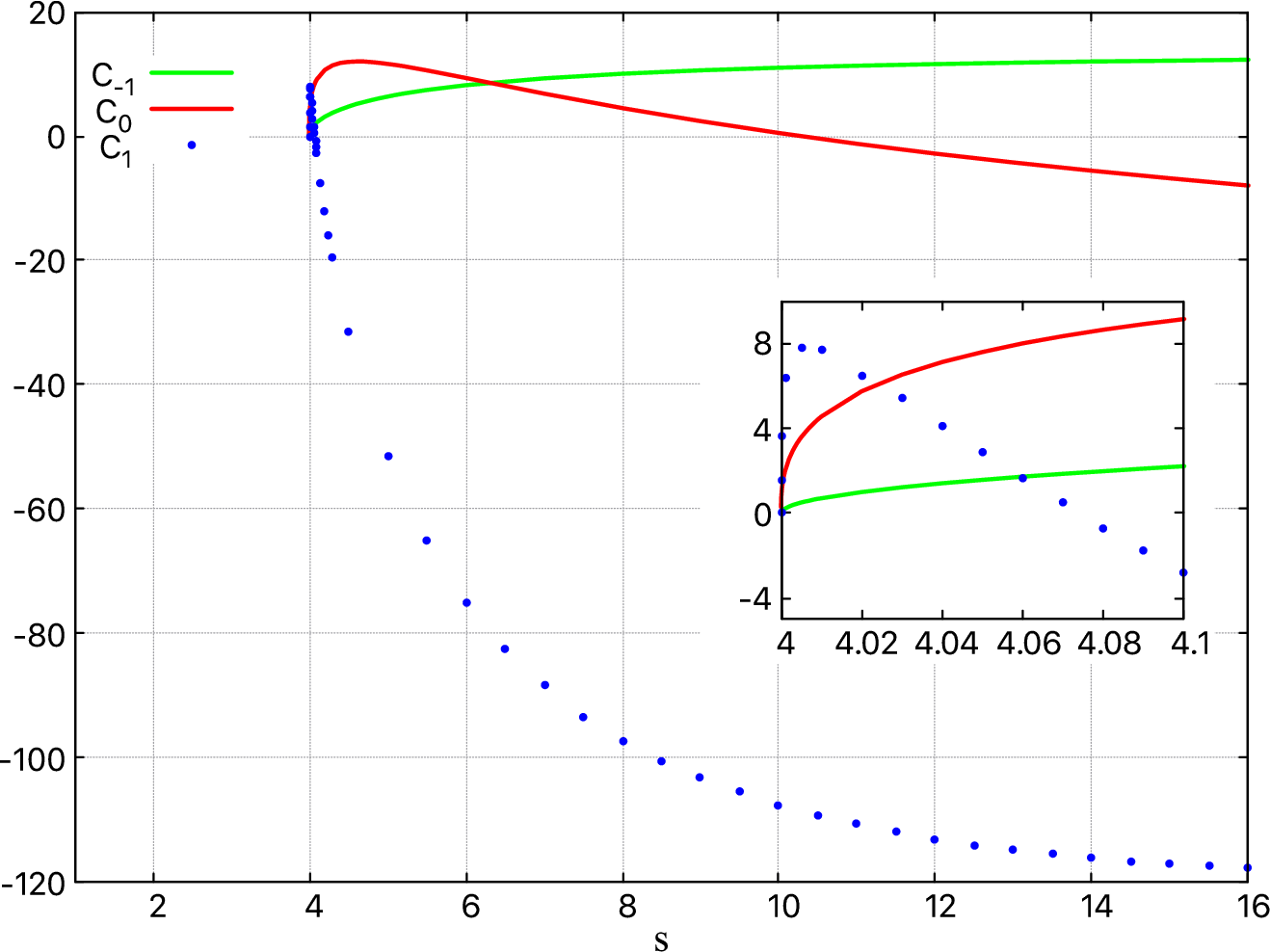}}
\end{center}
\caption{Expansion coefficients of Loop (II), $C_{-2}$ (dark-orange solid line), $C_{-1}$ (green solid line), $C_0$ (red solid line) and $C_1$ (blue dots) as a function of $s$. The coefficients around $s=4m^2$ are shown in the small inserted plot. (a) Real part; (b) Imaginary part. }
\label{fig:L-II}
\end{figure}

\begin{figure}[htb]
\begin{center}
\hspace*{0.3cm}
\subfloat[]{\includegraphics[width=0.49\linewidth]{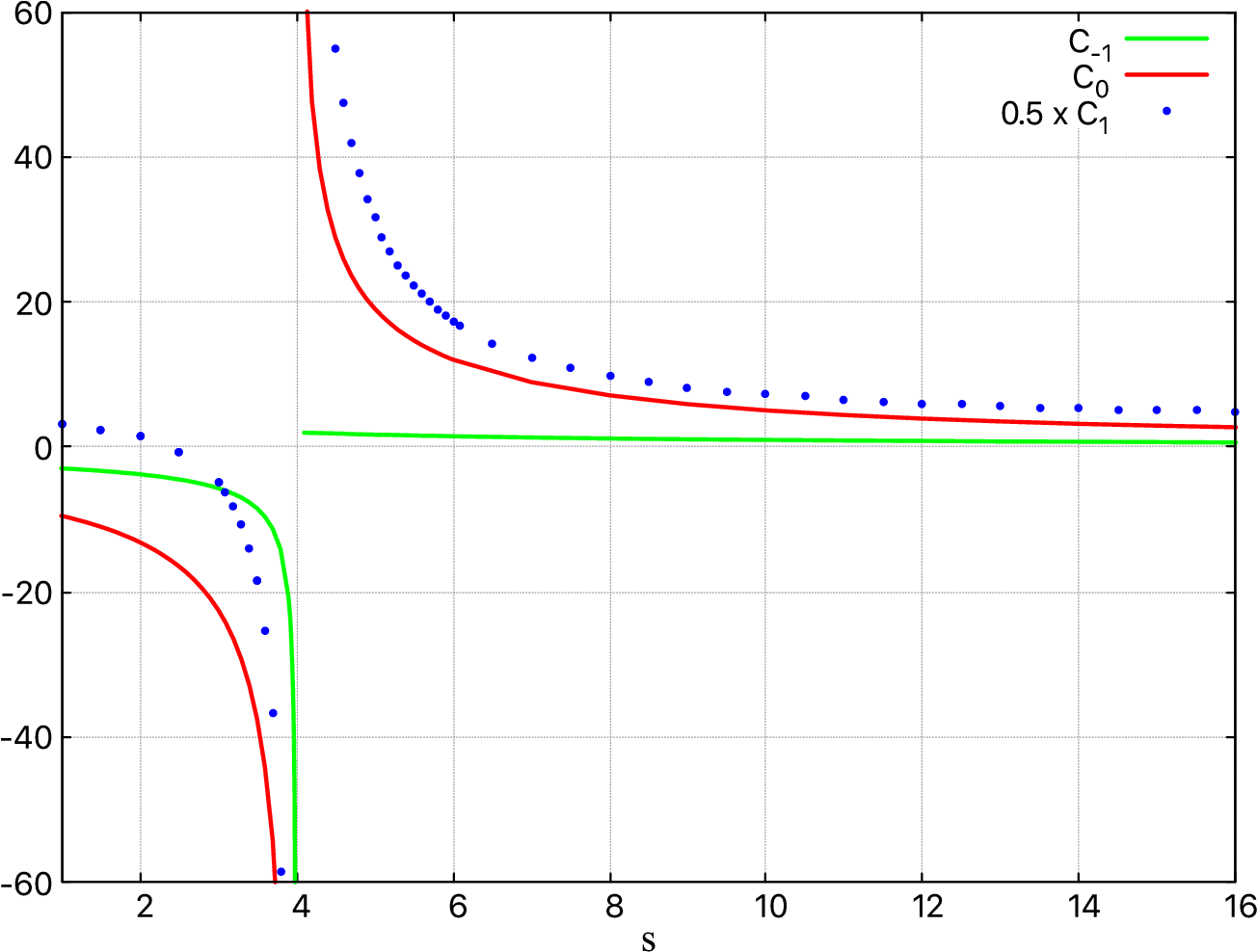}}
\hspace*{0.3cm}
\subfloat[]{\includegraphics[width=0.49\linewidth]{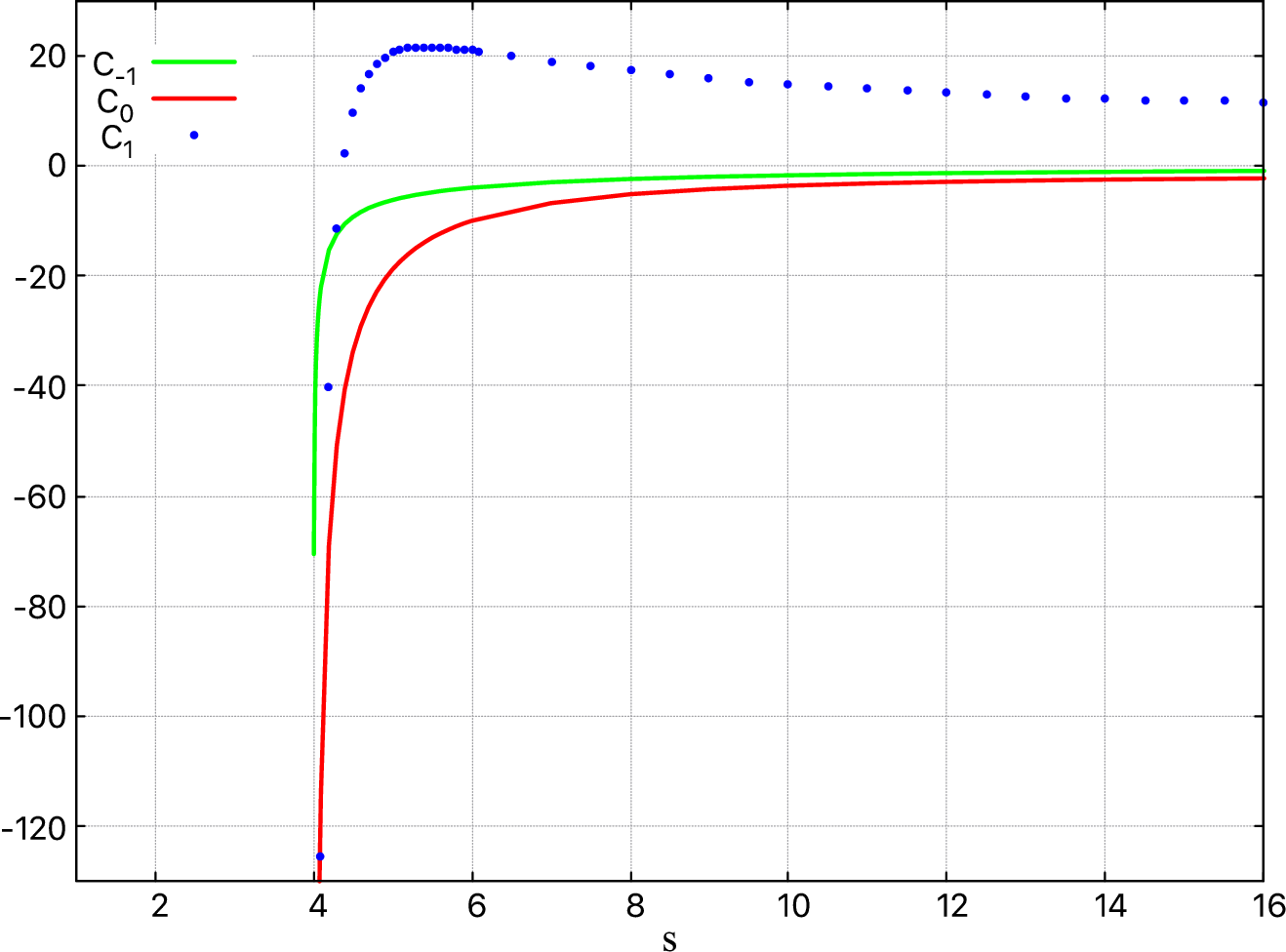}}
\end{center}
\caption{Expansion coefficients of Loop (III), $C_{-1}$ (green solid line), $C_0$ (red solid line) and $C_1$ (blue dots) as a function of $s$. (a) Real part, $C_1$ is multiplied by 0.5 to fit in the plot; (b) Imaginary part. }
\label{fig:L-III}
\end{figure}

\begin{figure}[htb]
\begin{center}
\hspace*{0.3cm}
\subfloat[]{\includegraphics[width=0.49\linewidth]{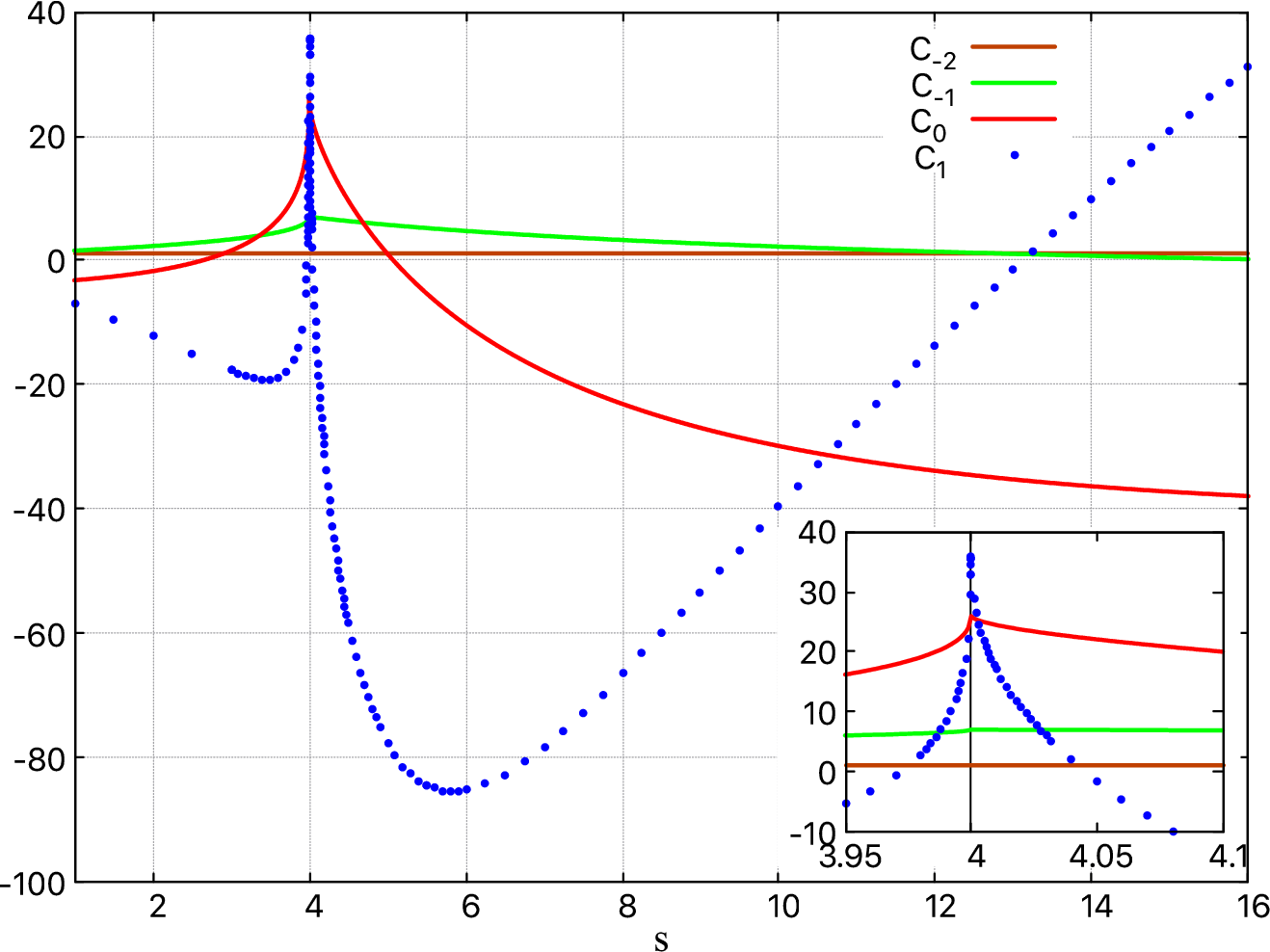}}
\hspace*{0.3cm}
\subfloat[]{\includegraphics[width=0.49\linewidth]{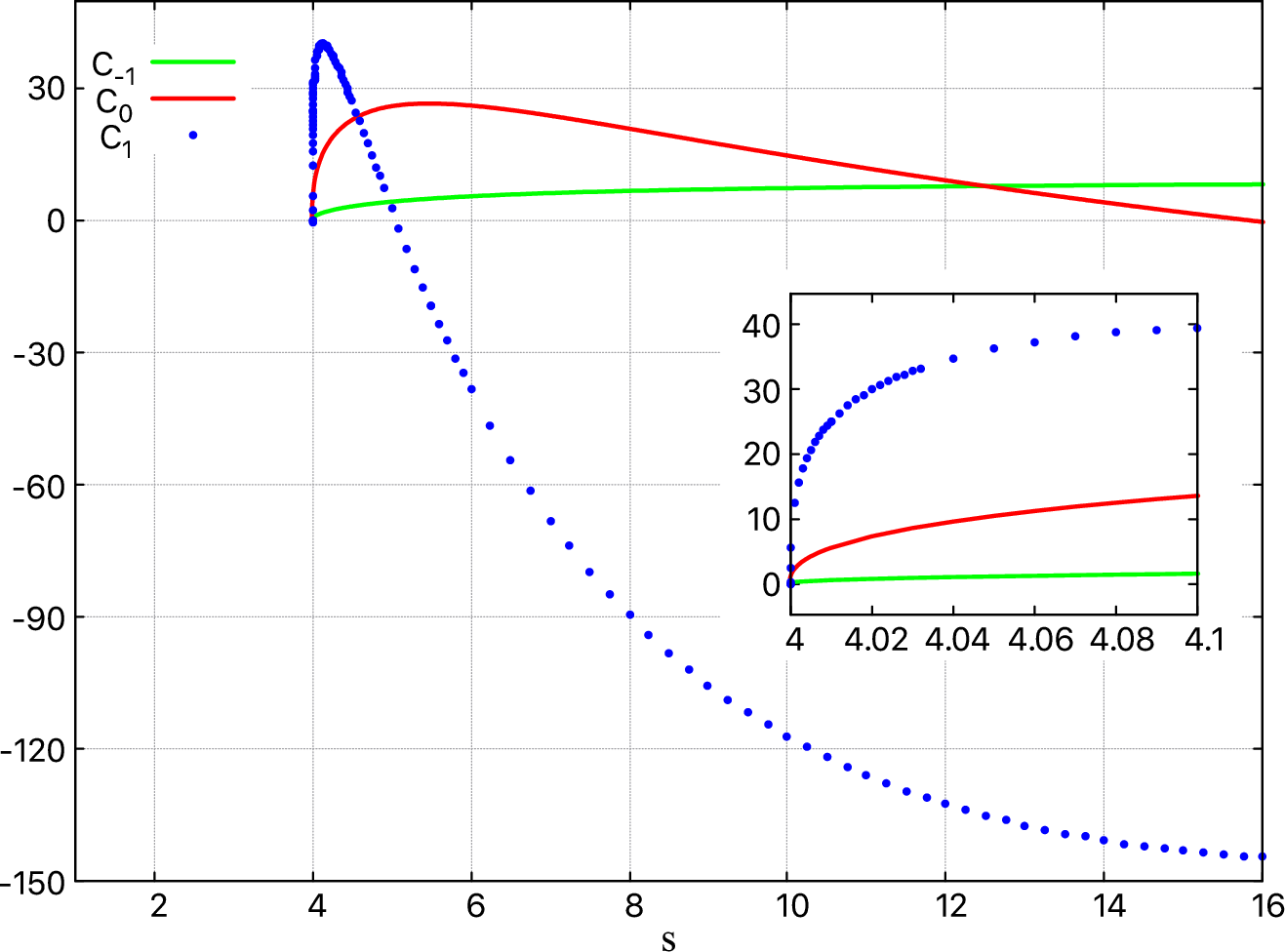}}
\end{center}
\caption{Expansion coefficients of Loop (IV), $C_{-2}$ (dark-orange solid line), $C_{-1}$ (green solid line), $C_0$ (red solid line) and $C_1$ (blue dots) as a function of $s$. The coefficients around $s=4m^2$ are shown in the small inserted plot. (a) Real part; (b) Imaginary part. }
\label{fig:L-IV}
\end{figure}

In Figures~\ref{fig:L-I}~-~\ref{fig:L-IV}, we present $C_{-2}, C_{-1}, C_0, C_1$ of Loop (I), (II), (III), and (IV)
as a function of $s$. Here we take the numerical value of $m=1$, so that the horizontal axis shows $s$ in units of $m^2$.
In these figures, the first three coefficients are shown in lines and the values of $C_1$ are
shown in dots.

In Fig.~\ref{fig:L-I}, we show Loop (I) and a moderate four-body threshold at $s=16m^2$ can be seen. 
The first three  coefficients are real. 
Except for Loop (I), we can observe a significant threshold at $s=4m^2$, though 
the behavior depends on each loop instance. In Loop (II) and (IV), the real
part shows a finite peak at $s=4m^2$, while it is divergent in Loop (III).
Any other possible threshold is numerically not very clear. 
In the figures for Loop (II), (III) and (IV), the behavior near $s=4m^2$ is shown
as the insertion.

There are choices for the numerical integration method as introduced in Section 1 
and in this paper we use the double-exponential formula~\cite{mori78}, or DE formula for short, 
to evaluate \siki{integfin}.
It  is an efficient method for the numerical integration whose integrand is 
an analytic function with singularities at the boundaries of the integration domain. 
This formula transforms the integration variable in $\int_0^1 f(x) dx$ to 
$x=\phi(t)=\frac{1}{2}(\rm{tanh}(\frac{\pi}{2}\rm{sinh}(t))+1)$. 
After the transformation, the trapezoidal rule is applied.
For multi-dimensional integrals, 
we use iterated integration, i.e., the DE formula is applied
in consecutive coordinate directions.
Depending on the behavior of the integrand, we select an adequate prescription shown in the Appendices.
With regard to the extrapolation, we use single extrapolation and double extrapolation as necessary.

Here, the $C$'s for Loop (I) are obtained by the double extrapolation 
and for the rest we use the single extrapolation in $\rho$
by use of  formulae resulting from the expansion in $\ep$
explained in the Appendix B.
As for the values of $\varepsilon$ and $\rho$ in the extrapolation, 
we typically use $\left(\frac{16}{15}\right)^{-\ell} (49\leq \ell \leq 59 )$ 
for $\ep$ and 
for $\rho$.
Meanwhile, for Loop (IV) we use much smaller $\rho$ as 
$\left(\frac{16}{15}\right)^{-\ell} (160\leq \ell \leq 179)$. 

For the first three terms, $C_{-2}$, $C_{-1}$, and $C_0$, the formulae 
in the previous section and the numerical results agree excellently 
including the regions close to the thresholds. 
Besides,  we confirm that all the coefficients obtained numerically  
also agree with those by pySecDec~\cite{borowka18a,borowka19,heinrich24} in the range
of $s$ shown in the figures. For example, when $s = 8 m^2$, 
the number of digits in agreement for the real part of Loop (I), (II), (III) and (IV) 
is  8, 4, 3 and 3, respectively. Similarly, for the imaginary part, 
we obtain 6, 4, 3, and 4 digits, respectively.
Particularly for $s=m^2$, the coefficients of Loop (I), (II) and (IV) 
show good agreement (10, 10 and 12 digits, respectively, 
for the $C_1$ term) with those by Laporta~\cite{laporta01}.

\section{Conclusion}

In this paper, we have studied the evaluation of
the coefficients of Laurent series in $\varepsilon$ 
for multi-loop integrals. 
Based on a simple SD,  we present an algorithm to obtain basic formulae 
which are suitable to be calculated analytically or numerically.
As a specific example, we select four 3-loop two-point 
diagrams whose $U$ function is  ``complete'' in four variables
\footnote{Note that, for the 2-loop two-point diagrams, besides those given as the product
of 1-loop diagrams, the $U$ function is  ``complete'' in three variables.
}.
Although the choice of diagrams is not essential, 
these examples conveniently demonstrate our approach
since  the $f$ function factored from the $U$ function  
is common in all regions.
We provide explicit analytic expressions for the coefficients of 
the UV divergent terms in Section 2,
and numerically compute all terms with reasonable accuracy in Section 3. 
These four diagrams are studied in the preceding works 
~\cite{martin23,martin22,borowka19} and our results 
reproduce them including the $s$-dependence.

Many multi-loop integrals have no such  ``complete'' property.
Also, when the masses of the internal lines are not the same,
the integrals for each region should be computed one by one.
However, the method introduced in the current paper can be applied to
many diagrams and to general mass configurations.
A preliminary study of some other 3-loop two-point functions for $N\le 6$ shows that 
they can be handled by the method discussed in the text.

The numerical method developed by the authors 
is able to calculate the coefficients of the Laurent series 
in the kinematical region both below and above the threshold.
The method shows good performance for the calculation of
the 3-loop integrals studied.


\section*{Acknowledgement}

We acknowledge the support by JSPS KAKENHI under Grant Numbers JP20K11858, \linebreak
JP20K03941 and JP21K03541, and by the National Science Foundation, Award Number
1126438 that funded initial work on numerical integration for the ParInt package.
We would like to express our deep gratitude to Dr. T. Kaneko 
for showing us with a clear explanation how to handle the SD method.

\section*{Appendix}
\appendix
\section{Separation of the UV singularity}
We briefly explain how the UV singularity is extracted in \siki{integfin}.
The basic tool is a subtraction to separate the $\ep$-pole.
Suppose one considers the integral
\begin{equation}
J=\int_0^1 dx\,\frac{1}{x^{1-\ep}}f(x,\ep)
\end{equation}
where $f(x,\ep)$ is regular at $x=0$. 
In the following, we write $f(x,\ep)$ as $f(x)$ for brevity.
Then  the subtraction is done as
\begin{equation}
J=\int_0^1 dx\,\frac{1}{x^{1-\ep}}[f(0) + (f(x)-f(0))]
 = \frac{1}{\ep}f(0) + \int_0^1 dx\,\frac{1}{x^{1-\ep}}[f(x)-f(0)].
\label{eq:subtract}
\end{equation}

If the integral is singular in two variables, the above procedure is
repeated:

\begin{eqnarray}
J&=& \int_0^1 dx\, \int_0^1 dy\, \frac{1}{x^{1-\ep}y^{1-\ep}} f(x,y) = \frac{1}{\ep^2} f(0,0)\nonumber \\
 & &+ \frac{1}{\ep} \int_0^1 dx\, \frac{1}{x^{1-\ep}} (f(x,0)-f(0,0)) + \frac{1}{\ep} \int_0^1 dy\, \frac{1}{y^{1-\ep}} (f(0,y)-f(0,0))\nonumber \\
 & &+ \int_0^1 dx\, \int_0^1 dy\, \frac{1}{x^{1-\ep}y^{1-\ep}}(f(x,y)-f(x,0)-f(0,y)+f(0,0)).
\label{eq:subtract2}
\end{eqnarray}

We also encounter an integral of the following type:
\begin{equation}
J=\int_0^1 dx\, \frac{1}{x^{2-\ep}}f(x),
\end{equation}
which needs analytic continuation for the singularity at $x=0$
\footnote{
We can use the numerical integration routine DQAGSE from QUADPACK~\cite{pi83,jocs11}, 
which performs a built-in extrapolation for the integral in Eq.~(A4).
}.
We use two methods to handle the integral.

\noindent Method-1: Expansion and subtraction
\begin{equation}
J= \frac{1}{-1+\ep}f(0) + \frac{1}{\ep}f'(0) 
+ \int_0^1 dx\, \frac{1}{x^{2-\ep}}(f(x) - f(0) - x f'(0))
\label{eq:method1}
\end{equation}

\noindent Method-2: Integration by parts
\[
J = \left[ \frac{x^{-1+\ep}}{-1+\ep} f(x) \right]_0^1 - \int_0^1 dx\, \frac{x^{-1+\ep}}{-1+\ep} f'(x)
\]
\begin{equation}
= \frac{1}{-1+\ep}f(1) + \frac{1}{(1-\ep)\ep}f'(0)
+ \frac{1}{1-\ep} \int_0^1 dx\, \frac{1}{x^{1-\ep}}(f'(x)-f'(0))
\label{eq:method2}
\end{equation}
Numerical evaluation by the two methods gives the same result.
From a pedagogical point of view, this is an explicit example
to demonstrate the uniqueness of the analytic continuation.
In the production stage of the computation in Section 3, the Method-2 is adapted.

\section{Expansion in  \texorpdfstring{$\boldsymbol\ep$}{epsilon}}
In \siki{subtract}, \siki{subtract2}, \siki{method1} and \siki{method2}, the UV singularity is explicitly 
shown as a factor $\ep^{-1}$ or $\ep^{-2}$ and
the integrals are no longer singular. Each term in the
right-hand side can be expanded as a series in $\ep$ up to the required order. 
Then we obtain an expression
to be computed analytically or numerically.

As an example, explicit computation of Loop (IV) is shown here.
We show the way to compute $I(klm)$ in \siki{expandone}  up to the $O(\ep)$.  
In the following equations, terms beyond $O(\ep)$ are omitted.

We use the following notations.
\begin{equation}
H= \frac{G^{-3\ep}}{f^{2-\ep}} ,\quad
H_a= \left. H\right|_{t=0} , \quad
H_b= \left. H\right|_{u=0} , \quad
H_0= \left. H\right|_{t=0,u=0} . 
\end{equation}
\begin{equation}
f=1+v+uv+tuv, \quad
f_a= \left. f\right|_{t=0}=1+v+uv , \ 
f_b= \left. f\right|_{u=0} , \ 
f_0= \left. f\right|_{t=0,u=0}=f_b=1+v ~, 
\end{equation}
and similar notation of the suffix-$a,b,0$ is used for  $G$ in Eq.~(20).
While $f$ is common in all regions, $G$ depends on the region.

The notation for the integral is as follows.
\[
\int d\Gamma_X= \int_0^1 dw_1 \int_0^1 dw_2 \int_0^1 dt \int_0^1 du \int_0^1 dv,\quad
\int d\Gamma_a= \int_0^1 dw_1 \int_0^1 dw_2 \int_0^1 du \int_0^1 dv ,
\] 
\begin{equation}
\int d\Gamma_b= \int_0^1 dw_1 \int_0^1 dw_2 \int_0^1 dt \int_0^1 dv , \quad
\int d\Gamma_0= \int_0^1 dw_1 \int_0^1 dw_2 \int_0^1 dv .
\end{equation}

\vspace{3mm}
\noindent \underline{$I(123)$}  \ \ 
In this integral, $t$- and $u$- divergences appear. Here $h=t$,  $G_a=G_0=m^2-s\bar{w}_1w_1$
and $G_b= m^2-s\bar{w}_1w_1+t(m^2-s\bar{w}_2w_2)$.

\begin{equation}
I(123)=  I(123)_0 +  I(123)_1 +  I(123)_2 +  I(123)_3 
\end{equation}
where $0, 1, 2, 3$ denote the integrals of $ H_0, H_a-H_0, H_b-H_0, H-H_a-H_b+H_0 $, respectively.

\[
I(123)_0= \frac{1}{2\ep^2}   
 \int d\Gamma_0 \frac{1}{f_0^2} \Bigl[
  1 + \ep \left( \log f_0 - 3 \log G_0 \right)
  + \frac{1}{2}\ep^2  \left( \log f_0 - 3 \log G_0 \right)^2
\]
\begin{equation}
  + \frac{1}{6} \ep^3 \left( \log f_0 - 3 \log G_0 \right)^3
  \Bigr] \,.
\label{eq:intonex}
\end{equation}
  
\[
 I(123)_1= \frac{1}{2\ep} \int d\Gamma_a \frac{1}{u} 
 \Bigl[ 
 \Bigl( \frac{1}{f_a^2}- \frac{1}{f_0^2} \Bigr)
\]
\[
 + \ep \Big(
 \frac{1}{f_a^2}(\log u+ \log f_a -3 \log G_0) 
 - \frac{1}{f_0^2}(\log u+ \log f_0 -3 \log G_0)
 \Big)
\]
\begin{equation}
 + \frac{1}{2} \ep^2 \Big(
 \frac{1}{f_a^2}(\log u+ \log f_a -3 \log G_0)^2 
 - \frac{1}{f_0^2}(\log u+ \log f_0 -3 \log G_0)^2
 \Big)
\Bigr] \,.
\label{eq:intoney}
\end{equation}

\[
I(123)_2 = \frac{1}{\ep} \int d\Gamma_b \frac{1}{t} \frac{1}{f_0^2}
  \Bigl[ (-3)\ep(\log G_b - \log G_0) 
\]
\begin{equation}
  + \ep^2 \Bigl( \frac{9}{2}((\log G_b)^2-(\log G_0)^2)
  -3(2\log t+\log f_0)(\log G_b - \log G_0) \Bigr)
  \Bigr] \,.
\label{eq:intonez}
\end{equation}

\[
I(123)_3= \int d\Gamma_X \frac{1}{tu} 
\Bigl[  \Bigl( \frac{1}{f^2}  -\frac{1}{f_a^2} \Bigr)
+ \ep \Bigl(
  \frac{1}{f^2}(2\log t + \log u + \log f - 3\log G) 
\] 
\begin{equation}
  - \frac{1}{f_a^2} (2\log t + \log u + \log f_a - 3\log G_0)
  + 3 \frac{1}{f_0^2}  (\log G_b - \log G_0)
  \Bigr)
\Bigr] \,.
\label{eq:intonew}
\end{equation}

\vspace{3mm}
\noindent \underline{$I(132), I(134)$} \ \ 
These two integrals have divergences from $t=0$ but they have no
divergence from $u=0$.  Here $h=tu, tuv$, where the factor $u$ cancels out $1/u$ in Eq.~(18), and
$G_a=m^2-s\bar{w}_1w_1$.

\begin{equation}
I(132, 134)= I(132, 134)_0 +  I(132, 134)_1
\end{equation}
where $0, 1$ indicate the integrals of $H_a, H-H_a $, respectively.

\begin{equation}
I(132, 134)_0= \frac{1}{2\ep}  
\int d\Gamma_a  \frac{(1, v)}{f_a^2} \Bigl[
1 + \ep ( \log u + \log f_a - 3 \log G_a )
+ \frac{1}{2} \ep^2 ( \log u + \log f_a - 3 \log G_a )^2
\Bigr] \,.
\label{eq:inttwox}
\end{equation}

\[
I(132, 134)_1=\int d\Gamma_X  
\frac{(1, v)}{t}
\Bigl[ \Bigl( \frac{1}{f^2} - \frac{1}{f_a^2} \Bigr) 
 + \ep \Bigl ( \frac{1}{f^2}(2\log t + \log u + \log f - 3\log G)
\]
\begin{equation}
 - \frac{1}{f_a^2}(2\log t + \log u + \log f_a - 3\log G_a) \Bigr) 
\Bigr] \,.
\label{eq:inttwoy}
\end{equation}

\vspace{3mm}
\noindent \underline{$I(312), I(341), I(314)$} \ \ 
These three integrals have no divergent term. Here $h=t^2u, t^2u^2v, t^2uv$.
\begin{equation}
I(312, 341, 314)= \int d\Gamma_X \frac{ (1, uv, v) }{f^2}
\left[ 
1 + \ep(2\log t + \log u + \log f - 3\log G)
\right]  \,.
\label{eq:intthrx}
\end{equation}

The integrals presented above for the $O(\ep^{-2})$, $O(\ep^{-1})$
and $O(1)$ terms are computed  analytically  to obtain \siki{analytic2j}.
The $O(\ep)$ terms are evaluated numerically
as shown in Section~3. 


\bibliographystyle{ptephy}
\bibliography{./bibptep}

\end{document}